\documentclass[conference]{IEEEtran}
\IEEEoverridecommandlockouts
\usepackage{cite}
\usepackage{amsmath,amssymb,amsfonts}
\usepackage{stfloats}
\usepackage{algorithmic}
\usepackage{graphicx}
\usepackage{textcomp}
\usepackage{xcolor}
\usepackage{booktabs}
\def\BibTeX{{\rm B\kern-.05em{\sc i\kern-.025em b}\kern-.08em
    T\kern-.1667em\lower.7ex\hbox{E}\kern-.125emX}}

\makeatletter
\newcommand{\linebreakand}{%
  \end{@IEEEauthorhalign}
  \hfill\mbox{}\par
  \mbox{}\hfill\begin{@IEEEauthorhalign}
}
\makeatother

\begin{document}

\title{Online Full ZVS Optimization for Modular Multi-Active Bridge Converter in MV PET\\
\thanks{\footnotesize This paper was supported by the Innovation and Technology Fund's Innovation and Technology Support Programme (ITP/018/24AP) and was supported by the National Natural Science Foundation of China under Grant 52577197.}
}

\author{
\IEEEauthorblockN{Haoyu Wang}
\IEEEauthorblockA{\textit{Department of Electrical Engineering} \\ \textit{Tsinghua University}\\ wanghaoy20@tsinghua.org.cn}
\and

\IEEEauthorblockN{Junwei Liu}
\IEEEauthorblockA{\textit{Dept. Electrical and Electronic Eng.} \\ \textit{The Hong Kong Polytechnic University}\\ junwei.jw.liu@polyu.edu.hk}
\and

\IEEEauthorblockN{Jialin Zheng}
\IEEEauthorblockA{\textit{Department of Electrical Engineering} \\
\textit{Tsinghua University}\\ zhengjl19@tsinghua.org.cn}
\linebreakand

\IEEEauthorblockN{Yangbin Zeng}
\IEEEauthorblockA{\textit{Department of Electrical Engineering} \\
\textit{South China University of Technology}\\ epybzeng@scut.edu.cn}
\and

\IEEEauthorblockN{Di Mou}
\IEEEauthorblockA{\textit{Department of Electrical Engineering} \\
\textit{Tsinghua University} \\ dimou428@mail.tsinghua.edu.cn}
\and

\IEEEauthorblockN{Zian Qin}
\IEEEauthorblockA{\textit{Department of Electrical Engineering} \\
\textit{Delft University of Technology} \\ z.qin-2@tudelft.nl}



}

\maketitle
\begin{abstract}
Multi-active bridge (MAB) converters, the core of the state-of-the-art medium-voltage power electronic transformers, can flexibly connect multiple DC ports among distributed DC grids and loads, but suffer from hard switching under conventional single phase-shift control, especially under unbalanced voltage conversion ratios and light load conditions. Although some offline methods manage to improve the efficiency through complex optimization structures, there lacks online optimization methods that are simple but effective due to the strong coupling among ports of the converter. By leveraging the time-domain model of the MAB converter under the multiple phase-shift modulation scheme, this paper simplifies the optimization process and proposes an online optimization method that can achieve full zero-voltage switching (ZVS) operation regardless of the load conditions. The proposed method has simple solutions with only voltage conversion ratios involved and can be implemented within a wide operation range without additional sensors or advanced controllers. A four-port MAB converter is constructed as the prototype. The simulation and experimental results have verified the feasibility and superiority of the proposed online strategy in achieving ZVS operation, dynamic response, and efficiency improvement.
\end{abstract}


\section{Introduction}
\begin{figure}[t]
\setlength{\abovecaptionskip}{-0.2cm}
\centerline{\includegraphics[width=9cm]{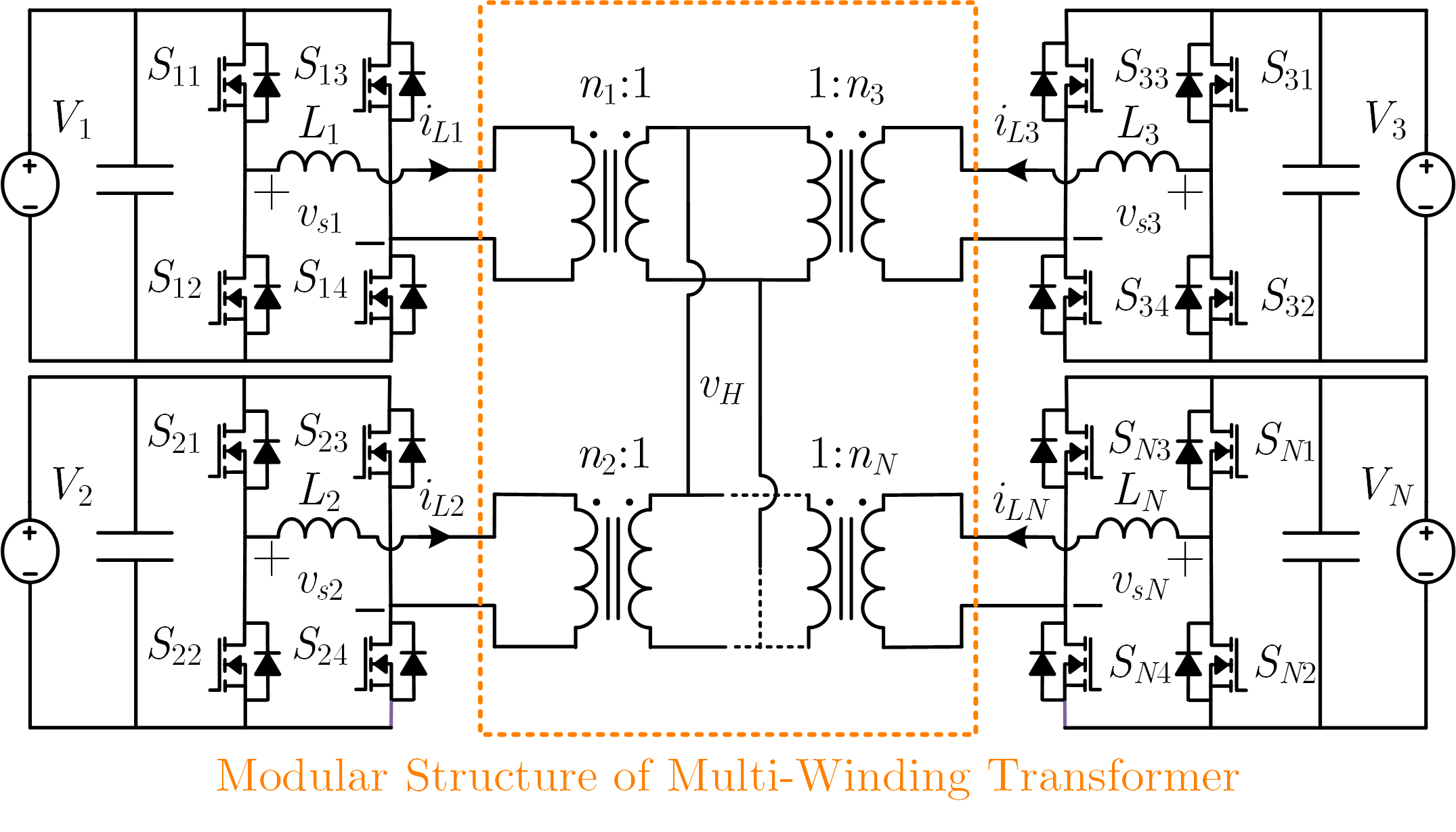}}
\caption{The typical configuration of a generalized modular MAB DC-DC converter.}
\label{fig:MAB}
\end{figure}

The rapid development of renewable energy sources exhibiting prominent features of sparse distribution and intermittent discontinuity has posed great challenges to local distribution grids. Vehicle-to-grid interaction where electric vehicles are largely involved is another important factor that urges the local grids to possess flexible capabilities of handling plug-and-play loads. Therefore, power electronic transformer (PET) is gradually playing a crucial role to flexibly distribute power flows among girds and loads with multiple voltage forms and levels, e.g., medium voltage (MV) \cite{a_q_huang, w_wen, k_li}. Specifically, multi-active bridge (MAB) DC-DC converters \cite{c_gu, l_costa}, the core stage of the state-of-the-art PET, can address power distribution concerns by providing multiple DC ports with a multi-winding transformer for flexible power conversion. The multi-winding transformer can be further replaced by multiple transformers, leading to modular MAB converters that are capable of flexible power and voltage ratings \cite{p_zumel, l_ortega}, as shown in Fig. \ref{fig:MAB}. However, MAB converters adopting traditional single phase-shift (SPS) control suffer from hard switching and low efficiency under unbalanced voltage conversion ratios and light load conditions \cite{p_purgat, h_wang}. 

Some offline methods in the literature have managed to increase the efficiency of the multiport converter, which typically originate from the phase-shift optimization of two-port dual active bridge (DAB) converters \cite{b_zhao, n_hou, b_zhao_2, a_tong, b_zhao_3}, but can only be implemented offline due to the strong coupling among different ports. Specifically, an offline particle swarm optimization method to achieve ZVS and minimized root-mean-square (RMS) currents for a triple active bridge converter was proposed in \cite{j_li}. A look-up-table-based power-loss optimization method has been proposed for generalized modular MAB converters in \cite{h_wang_2}, but the ZVS operation was not an optimization objective and the efficiency improvement was relatively low in this case. By utilizing multiple control degrees of freedom, \cite{h_wang_3} achieved full ZVS operation and minimized RMS currents for modular MAB converters with any given ports, but significantly relied on a look-up table for global optimization. These offline methods are potentially prohibitive for smooth dynamics. Therefore, reliable online designs are needed to avoid poor dynamic performance while ensuring ZVS and strengthening efficiency. 

Existing online optimization methods are typically based on simplified mathematical models or closed-loop controllers \cite{d_mou_2, h_wang_5}. Particularly, \cite{d_mou} proposed a time-division multiplexing modulation method that flexibly changes the structure of port bridges in different ports between half bridges and full bridges, thus reducing conducting losses. However, it can only handle light loads under certain working conditions and lose the optimization capabilities within the full range. Multiple PI controllers have been used in \cite{o_hebala} to track the minimum point of the overall RMS currents, but the method may experience oscillations and poor dynamics. Besides, a reactive-power optimal control strategy was proposed in \cite{h_wang_4} to eliminate the fundamental reactive power in the MAB converter, thus leading to reduced RMS currents. However, the dual-loop control requires additional high-frequency current sensors and conditioning circuits for fundamental component extraction.

This paper proposes a simple online solution for MAB converters to achieve full ZVS within a wide operation range regardless of the load conditions, which ensures great dynamic responses to load changes. Without the need for an accurate model and parameters, only DC port voltages that are originally sampled in closed-loop PI  control are involved in the design, which can be easily implemented in practice. 

The rest of the paper is organized as follows: Section \ref{sec:modeling} introduces the operation principle and the modeling of the converter under multi-phase-shift modulation, the online ZVS optimization approach is given out in Section \ref{sec:optimization}, Section \ref{sec:verification} verifies the effectiveness of the proposed method by simulation and experimental results, and Section \ref{sec:conclusion} concludes the paper. 

\section{Operation and Modeling of MAB Converters}\label{sec:modeling}
\subsection{Operation Principles under Phase-Shift Modulation}
The typical configuration of a generalized MAB converter is shown in Fig. \ref{fig:MAB}. The generalized converter has $N$ ports with each port having an index $i$ ($i=1, 2, \cdots, N$). Each port $i$ consists of a full bridge of power semiconductors $S_{i1-i4}$ (e.g., SiC MOSFETS) and their anti-parallel diodes $D_{i1-i4}$. The DC voltage of each port is denoted as $V_{i}$ and the switched-node voltage of the full bridge is denoted as $v_{si}$. The capacitance at the DC side is $C_{i}$. There is a multi-winding transformer in the converter that can be considered as multiple modular transformers connected by a high-frequency link. The high-frequency link only exists in modular MAB converters but can be used to model the generalized MAB converter in this case. With the modular structure, $L_{i}$ represents the total leakage inductance of the transformer with a turns ratio of $n_{i}:1$ in port $i$. The voltage of the high-frequency link is $v_{H}$.

\begin{figure}[t]
\centerline{\includegraphics[width=9cm]{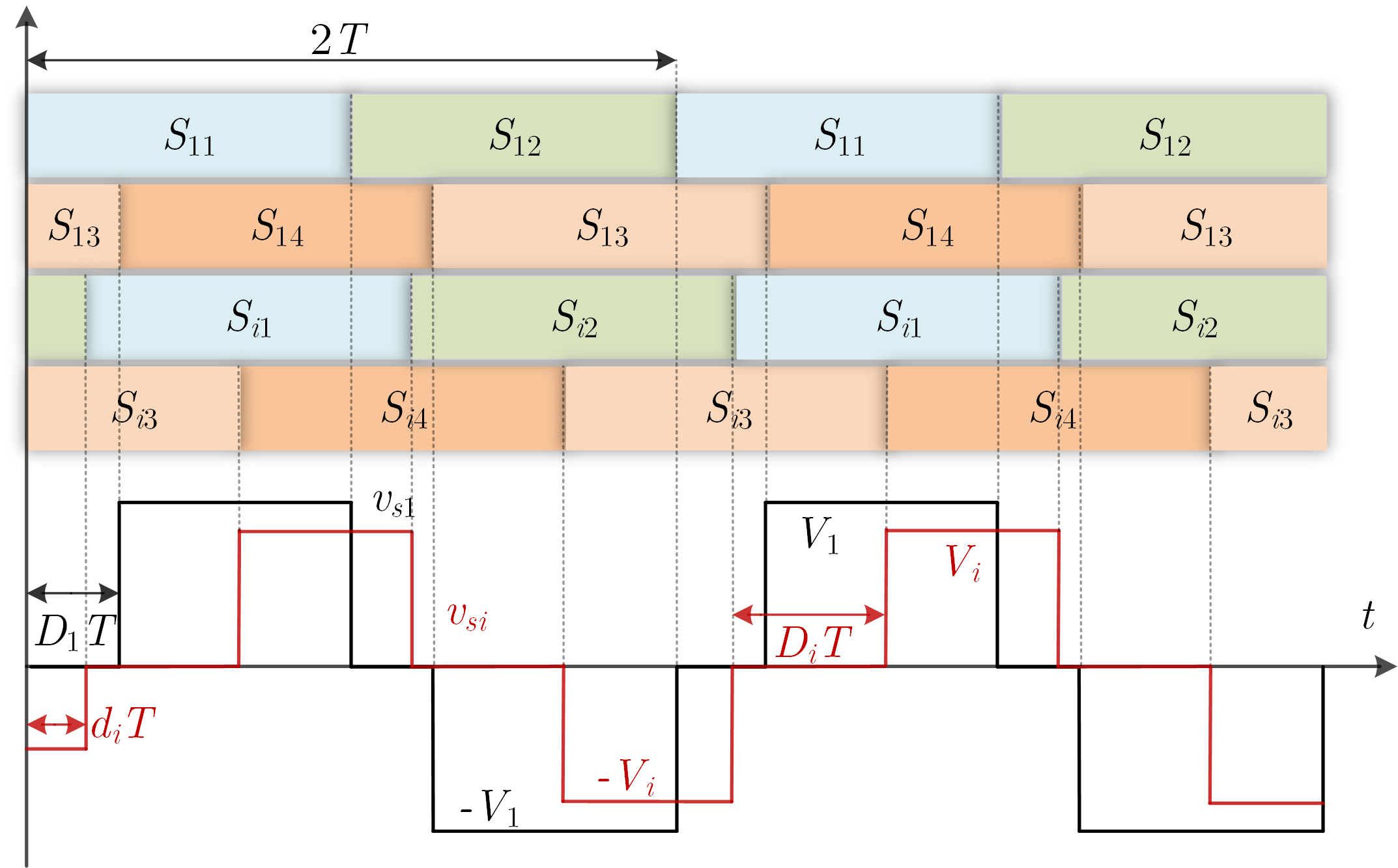}}
\caption{The switching orders of power semiconductors and phase shifts among ports.}
\label{fig:switch}
\end{figure}

Under traditional phase-shift modulation, the switching orders of active power semiconductors are square waves with $50\%$ duty cycles, as shown in Fig. \ref{fig:switch}. The switch pairs in the same arm (e.g., $S_{i1}$ and $S_{i2}$) are complimentary but the diagonal pairs (e.g., $S_{i1}$ and $S_{i4}$) can have phase shifts. Therefore, an outer phase-shift ratio $d_{i}$ ($-0.5\leq d_{i}\leq 0.5$) and an inner phase-shift ratio $D_{i}$ ($0\leq D_{i}< 1$) are independently obtained in a period $2T$, resulting in three-level quasi-square $v_{si}$ that has certain phase shift corresponding to the reference port. Note that $d_{1}$ is defined as zero for reference and all other $d_{i}$ are usually controlled by closed loops while all $D_{i}$ can be independently given according to some manners. In typical designs, all $d_{i}$ are used to regulate DC voltages or meet the power demands and all $D_{i}$ provide a large number of control degrees of freedom for optimization (e.g., efficiency), leading to a versatile multi-phase-shift modulation scheme.

\subsection{Mathematical Model with Multiple Phase Shifts}
Under multi-phase-shift modulation, any switching action of the converter will result in a new operation mode, thus leading to piecewise expressions of inductor currents at each operation mode, which takes a lot of expert efforts to manually analyze each operation mode and derive them. Therefore, it is significantly challenging to mathematically model an MAB converter in time domain. However, the time-domain inductor currents have time information and maintain great modeling accuracy, which is essential to identify ZVS conditions at switching instants. To address this problem, \cite{h_wang_3} has proposed a decomposition-and-superposition method that uses continuous waveforms to model the MAB converter.

\begin{figure}[t]
\setlength{\abovecaptionskip}{-0.2cm}
\centerline{\includegraphics[width=9cm]{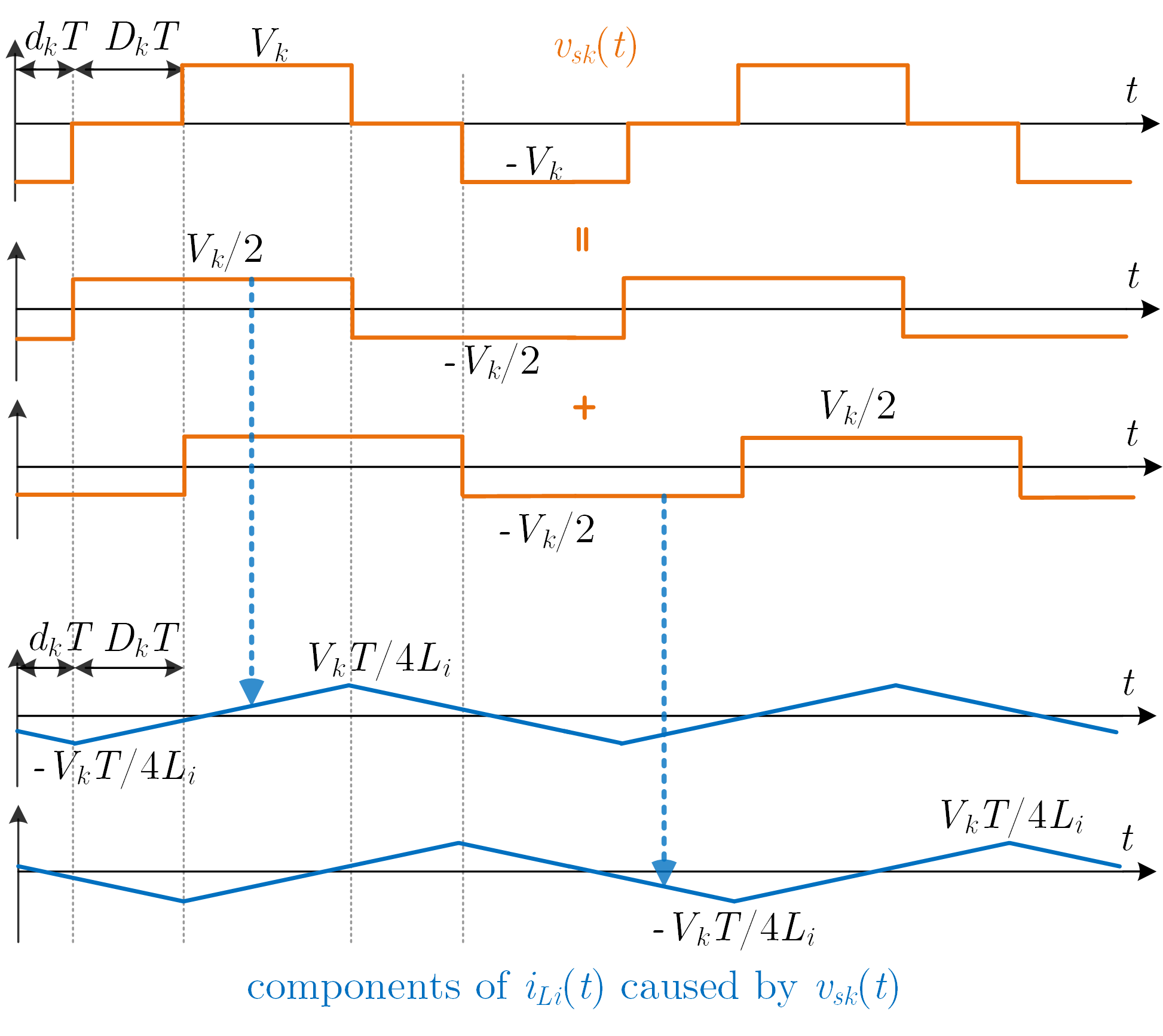}}
\caption{The decomposition of $v_{sk}$ into two square waveforms and the corresponding triangular components of $i_{Li}$.}
\label{fig:decomp_superp}
\end{figure} 

To begin, denote port 1 as reference and define the DC voltage conversion ratios for each port as:  
\begin{equation} \label{}
M_{i}=\frac{n_{1}V_{i}}{n_{i}V_{1}}.
\end{equation}

In addition, define the inductor coefficients of each port as:
\begin{equation} \label{}
l_{i}=\frac{\frac{n_{i}^{2}}{L_{i}}}{\sum_{k=1}^{N}\frac{n_{k}^{2}}{L_{k}}}, \quad \text{where} \quad \sum_{i=1}^{N}l_{i}=1.
\end{equation}

Suppose there is a virtual or physical high-frequency link. Its voltage $v_{H}$ can be expressed by a weighted sum of all $v_{si}$.  
\begin{equation} \label{}
v_{H}(t)=\sum_{i=1}^{N}\frac{l_{i}}{n_{i}}v_{si}(t).
\end{equation}

The inductor current $i_{Li}$ in port $i$ is impacted by both $v_{H}$ and $v_{si}$. Therefore, $i_{Li}$ can be expressed as:
\begin{equation} \label{eq:inductor_current_ode}
L_{i}\frac{i_{Li}(t)}{dt} =v_{si}(t)-n_{i}\sum_{k=1}^{N}\frac{l_{k}}{n_{k}}v_{sk}(t).
\end{equation}

\begin{figure*}[b]
\begin{equation} \label{eq:current_expression}
i_{Li}(t)=\frac{n_{i}}{n_{1}}\frac{V_{1}T}{2L_{i}}\left[\sum_{k=1}^{N}l_{k}M_{k}-M_{i}+l_{i}\left(\left\vert \frac{t}{T}-d_{i}\right\vert+\left\vert \frac{t}{T}-d_{i}-D_{i}\right\vert\right)-\sum_{k=1}^{N}l_{k}M_{k}\left(\left\vert \frac{t}{T}-d_{k}\right\vert+\left\vert \frac{t}{T}-d_{k}-D_{k}\right\vert\right) \right].
\end{equation}
\end{figure*}

\begin{figure*}[b]
\begin{equation} \label{eq:inductor_current_instant}
 \left\{
 \begin{array}{l}
 i_{Li}(t_{i1})=\frac{n_{i}}{n_{1}}\frac{V_{1}T}{2L_{i}} \left[\sum_{k=1,k\neq i}^{N}l_{k}M_{k}\left(1-D_{k}\right) 
 -(1-l_{i})M_{i}\left(1-D_{i}\right)-\sum_{k=1}^{N}2l_{k}M_{k}F_{1}(i,k)  \right], \\
 i_{Li}(t_{i2})=\frac{n_{i}}{n_{1}}\frac{V_{1}T}{2L_{i}} \left[\sum_{k=1,k\neq i}^{N}l_{k}M_{k}\left(1-D_{k}\right) 
 -(1-l_{i})M_{i}\left(1-D_{i}\right)-\sum_{k=1}^{N}2l_{k}M_{k}F_{2}(i,k)  \right]. 
 \end{array}
 \right.
\end{equation}
\end{figure*}

Because \eqref{eq:inductor_current_ode} is a linear equation, all $v_{sk}$ are supposed to have corresponding effects on $i_{Li}$. It is possible to analyze the specific impact of certain $v_{sk}$ on $i_{Li}$ with a designated weight ($k$ can be different from or the same as $i$) and then add them all together. Following this path, a further simplification has been made: a 3-level $v_{sk}$ is decomposed into two 2-level square waveforms with $50\%$ duty cycles and half the magnitude, as shown in Fig. \ref{fig:decomp_superp}. The two 2-level square waveforms have different phases corresponding to the outer and inner phases of port $i$, and will then result in two triangular components of $i_{Li}$ in steady state. The triangular inductor components also maintain the phase-shift information and their specific values are shown in Fig. \ref{fig:decomp_superp}. With all inductor components and known weights in \eqref{eq:inductor_current_ode}, the overall $i_{Li}$ can be expressed by \eqref{eq:current_expression}.

\section{Online Full ZVS Optimization Design}\label{sec:optimization}

\subsection{ZVS Operation Under Multi-Phase-Shift Modulation}

MAB converters under phase-shift modulation have inherent ZVS characteristics. The ZVS operation is determined by the inductor current at the switching instants. Taking into account the symmetry due to the $50\%$ duty cycle modulation, the switching pairs (e.g., $S_{i1}$ and $S_{i2}$) will maintain or lose ZVS operation simultaneously. Therefore, it is only essential to analyze the switching-instant inductor currents of $S_{i1}$ and $S_{i4}$. Define $t_{i1}=d_{i}T$ and $t_{i2}=\left(d_{i}+D_{i}\right)T$ as the switching instants of $S_{i1}$ and $S_{i4}$, respectively. According to \eqref{eq:current_expression}, the inductor currents in port $i$ at these instants can be expressed by 
\eqref{eq:inductor_current_instant}, where $F_{1}(i,k)$ and $F_{2}(i,k)$ are:

\begin{align}
\label{eq:embedded_function1}
F_{1}(i,k)=
\begin{cases}
d_{i}-d_{k},            & d_{i} \ge d_{k}, \\
d_{k}-d_{i}-D_{i},      & d_{k} \ge d_{i}+D_{i}, \\
0,                     & \text{else}.
\end{cases}
\\[6pt]
F_{2}(i,k)=
\begin{cases}
d_{k}-d_{i},            & d_{i} \le d_{k}, \\
d_{i}-D_{i}-d_{k},      & d_{i} \ge d_{k}+D_{i}, \\
0,                     & \text{else}.
\end{cases}
\end{align}

Ideally, the ZVS operation of $S_{i1}$ and $S_{i4}$ can be achieved if $i_{Li}$ is negative at $t_{i1}$ and $t_{i2}$ when the positive directions are shown in Fig. \ref{fig:MAB}, thus leading to the ZVS operation of $S_{i2}$ and $S_{i3}$. It is notable that $F_{1}(i,k)$ and $F_{2}(i,k)$ are always non-negative. Therefore, the full ZVS term $T_{i}$ of port $i$, where no outer phase shifts indicating powers are involved, can be defined as: 
\begin{equation}\label{eq:full_ZVS_term}
    T_{i}=\sum_{k=1,k\neq i}^{N}l_{k}M_{k}\left(1-D_{k}\right)-(1-l_{i})M_{i}\left(1-D_{i}\right).
\end{equation}

\subsection{Online Full ZVS Optimization}

According to \eqref{eq:inductor_current_instant}, when $T_{i} \leq 0$, $i_{Li}(t_{i1})$ and $i_{Li}(t_{i2})$ are non-positive as well, thus leading to ideal ZVS operation regardless of the working conditions. Note that all $D_{i}$ can be manually adjusted, which makes it possible to achieve ZVS operation within the full operation range. To make sure all switches achieve ZVS, for any port $i$, there must be $T_{i} \leq 0$. However, there exists $\sum_{i=1}^{N}l_{i}T_{i}=0$. Therefore, the full ZVS operation of the whole $N$-port converter can only be achieved when $T_{i}=0$, which leads to \eqref{eq:full_ZVS_optimization}.

\begin{figure}[t]
\centerline{\includegraphics[width=9cm]{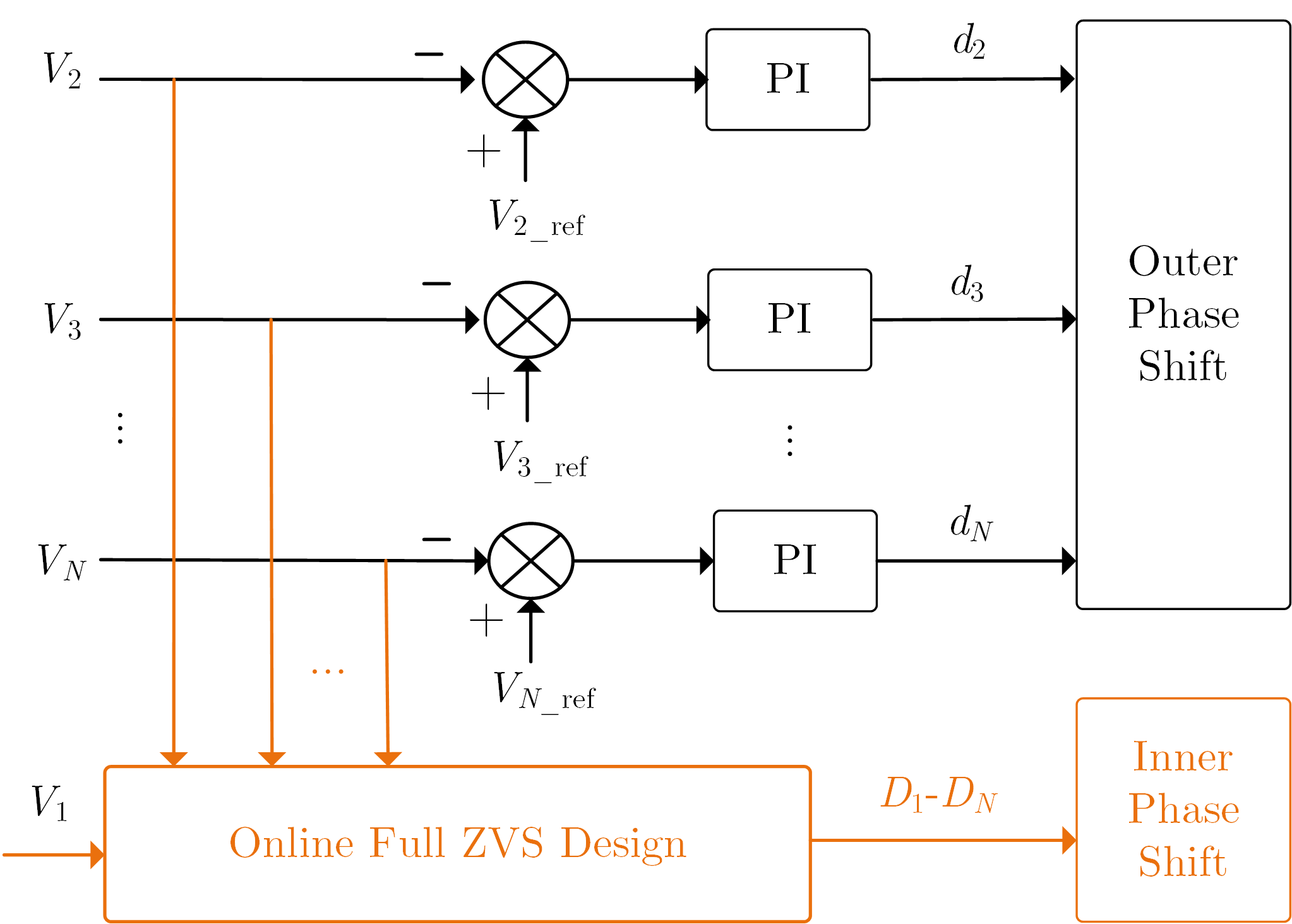}}
\caption{The overall control scheme with online full ZVS optimization design.}
\label{fig:control}
\end{figure} 

\begin{equation}\label{eq:full_ZVS_optimization}
\begin{aligned}
  & \left[\begin{matrix}
       (1-l_{1})M_{1} & -l_{2}M_{2} & \cdots & -l_{N}M_{N} \\
       -l_{1}M_{1} & (1-l_{2})M_{2} & \cdots & -l_{N}M_{N} \\
       \vdots & \vdots & \ddots & \vdots \\
       -l_{1}M_{1} & -l_{2}M_{2} & \cdots & (1-l_{N})M_{N}
       \end{matrix}\right]
    \left[\begin{matrix}
       D_{1} \\
       D_{2} \\
       \vdots \\
       D_{N}
    \end{matrix}\right]
       \\
  & =
    \left[\begin{matrix}
       M_{1}-\sum_{i=1}^{N}l_{i}M_{i} \\
       M_{2}-\sum_{i=1}^{N}l_{i}M_{i} \\
       \vdots \\
       M_{N}-\sum_{i=1}^{N}l_{i}M_{i}
    \end{matrix}\right].    
\end{aligned}
\end{equation}

The general solution of \eqref{eq:full_ZVS_optimization} can be obtained as:

\begin{equation}\label{eq:full_ZVS_solution}
\begin{aligned}
      \left[\begin{matrix}
       D_{1} \\
       D_{2} \\
       \vdots \\
       D_{N}
    \end{matrix}\right]
    =
    \left[\begin{matrix}
       1-\frac{\lambda}{M_{1}} \\
       1-\frac{\lambda}{M_{2}} \\
       \vdots \\
       1-\frac{\lambda}{M_{N}}
    \end{matrix}\right]
    , \quad \lambda \in \left(0, \text{min}\left\{M_{1}, M_{2}, \cdots, M_{N} \right\} \right].
\end{aligned}
\end{equation}

To ensure maximum power delivery capacities, $D_{i}$ should be as small as possible so that the switched-node voltages are more likely to be two-level instead of three-level. Therefore, an effective simplification has been adopted in this paper where $\lambda = \text{min}\left\{M_{1}, M_{2}, \cdots, M_{N} \right\}$ so that all $D_{i}$ are minimized while maintaining the full ZVS capability. Suppose that port $j$ has the minimum voltage conversion ratio $M_j$ in real-time applications, \eqref{eq:full_ZVS_solution} is converted into:

\begin{equation}\label{eq:full_ZVS_solution_sim}
\begin{aligned}
      \left[\begin{matrix}
       D_{1} \\
       D_{2} \\
       \vdots \\
       D_{N}
    \end{matrix}\right]
    =
    \left[\begin{matrix}
       1-\frac{M_{j}}{M_{1}} \\
       1-\frac{M_{j}}{M_{2}} \\
       \vdots \\
       1-\frac{M_{j}}{M_{N}}
    \end{matrix}\right]
    , \quad M_{j}=\text{min}\left\{M_{1}, M_{2}, \cdots, M_{N} \right\}.
\end{aligned}
\end{equation}

\begin{table}[t]
\centering
\caption{Main Parameters of the 4-Port MAB Converter}
\begin{tabular}{@{}lcc@{}}
\toprule
\textbf{Parameter} & \textbf{Symbol} & \textbf{Value/Type} \\ \midrule
Port 1 DC voltage & $V_1$ & 400 V \\ 
Port 2 DC voltage & $V_2$ & 500 V \\ 
Port 3 DC voltage & $V_3$ & 200 V \\ 
Port 4 DC voltage & $V_4$ & 300 V \\ 
Port 1 leakage inductance & $L_1$ & 15 $\mu$H \\
Port 2 leakage inductance & $L_2$ & 20 $\mu$H \\
Port 3 leakage inductance & $L_3$ & 8 $\mu$H \\
Port 4 leakage inductance & $L_4$ & 50 $\mu$H \\
Port 2 DC load & $R_2$ & 500 $\Omega$ \\
Port 3 DC load & $R_3$ & 100 $\Omega$ \\
Port 4 DC load & $R_4$ & Electronic load \\
Port 1 turns ratio & $n_1 : 1$ & 1:1 \\
Port 2 turns ratio & $n_2 : 1$ & 1:1 \\
Port 3 turns ratio & $n_3 : 1$ & 0.5:1 \\
Port 4 turns ratio & $n_4 : 1$ & 1:1 \\
Switching frequency & $f_s$ & 50 kHz \\ 
\bottomrule
\end{tabular}
\label{tab:parameter}
\end{table}

DC port voltages are usually sampled for closed-loop voltage or power regulation in conventional SPS control. Therefore, the voltage conversion ratios can be acquired in real-time sampling and require no additional sensors.  To conclude, \eqref{eq:full_ZVS_solution_sim} is an online optimization design with no additional cost. It does not require precise model parameters, such as the inductances $L_{i}$ of all ports and parasitic stray parameters, which is conducive to the full ZVS operation of any given generalized MAB converters without additional expert efforts. 

The overall control diagram of the online full ZVS design is shown in Fig. \ref{fig:control}. Note that port 1 is the reference port and handles power according to other ports, which means that there is no control loop. PI controllers are maintained to obtain outer phase shifts for the power regulation of all other ports. The ZVS design is achieved by using the sampled DC voltages and real-time calculation. The online optimization together with the closed-loop control forms the overall full ZVS design.

\section{Simulation and Experimental Verification}\label{sec:verification}

In order to verify the effectiveness of the proposed online full ZVS optimization strategy, a 4-port MAB converter has been constructed for simulations and experiments. The detailed parameters of the converter are shown in Table \ref{tab:parameter}. 

\begin{figure*}[ht]
\setlength{\abovecaptionskip}{-0.2cm}
\centerline{\includegraphics[width=18cm]{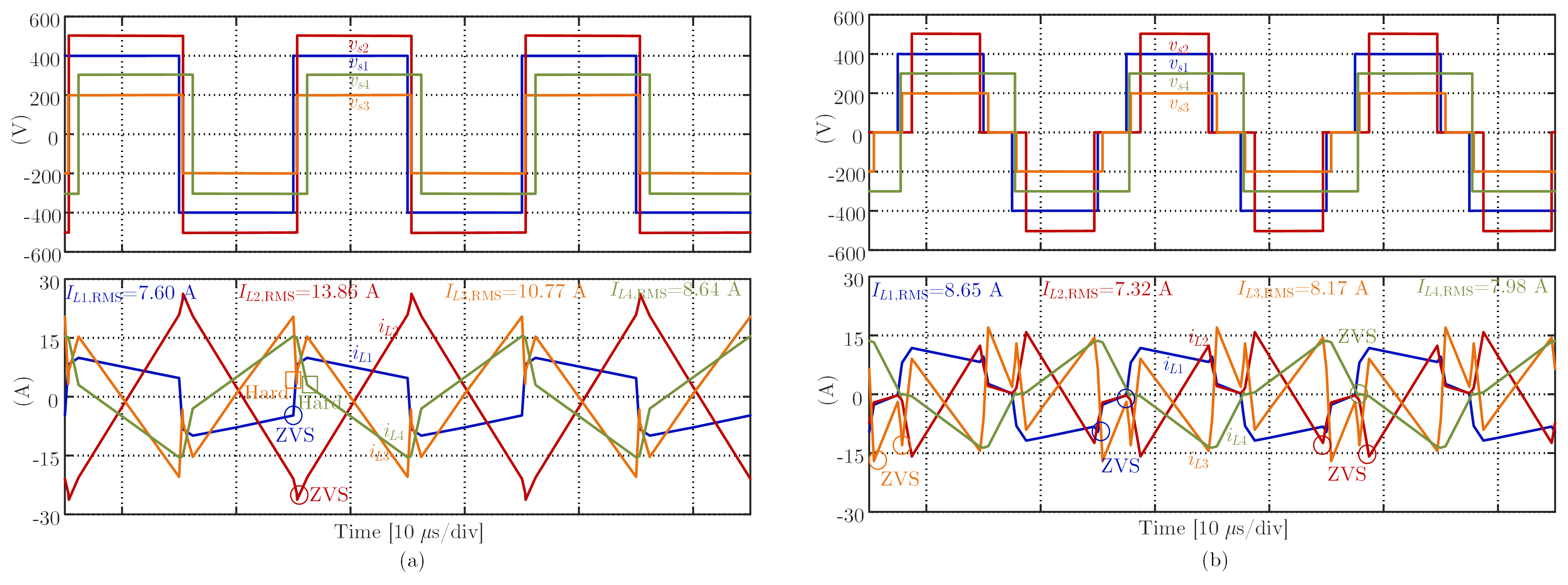}}
\caption{Simulated steady-state high-frequency waveforms when $P_4$ = 2 kW. (a) Under SPS control. (b) Under online ZVS design.}
\label{fig:simulation_1}
\end{figure*}

Specifically, port 1 is defined as the reference port and is connected to a 400 V DC voltage source ($V_1$ = 400 V) while ports 2-4 are connected to DC loads. The switching frequency is selected as 50 kHz and the turns ratios of the transformers are $n_1: n_2: n_3: n_4$ = 1: 1: 0.5: 1. The DC reference voltages of ports 2-4 are set as $V_2$ = 500 V, $V_3$ = 200 V and $V_4$ = 300 V, which leads to the voltage conversion ratios as $M_1: M_2: M_3: M_4$ = 1: 1.25: 1: 0.75. The inductances of the ports are $L_1$ = 15 $\mu$H, $L_2$ = 20 $\mu$H, $L_3$ = 8 $\mu$H, and $L_4$ = 15 $\mu$H. The loads of ports 2 and 3 are fixed as 500 $\Omega$ and 100 $\Omega$, leading to their handled powers $P_2$ = 0.4 kW and $P_3$ = 0.5 kW, while port 4 is connected to an electronic load so that the power handled by port 4 can be flexibly adjusted for different working conditions. For example, $P_4$ is set as 0.4 kW and 2 kW under two working conditions for further steady-state and dynamic tests in both simulations and experiments.

\begin{figure}[t]
\setlength{\abovecaptionskip}{-0.2cm}
\centerline{\includegraphics[width=7cm]{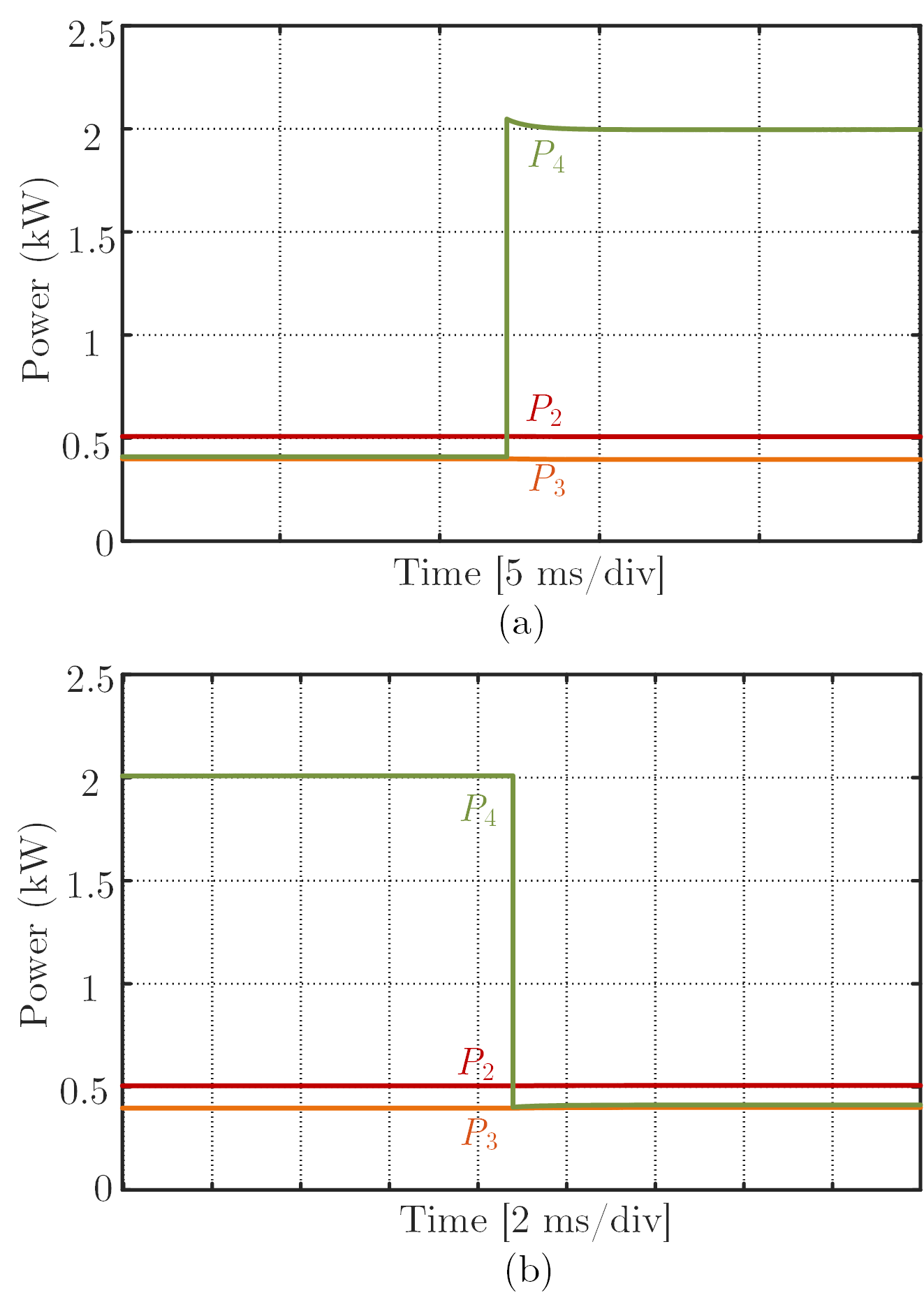}}
\caption{The simulated dynamics of the online ZVS design between two conditions. (a) When $P_4$ changes from 0.4 kW to 2 kW. (b) When $P_4$ changes from 2 kW to 0.4 kW.} 
\label{fig:simulation_2}
\end{figure}

\subsection{Simulation Results}\label{subsec:simulation}

\begin{figure}[t]
\setlength{\abovecaptionskip}{-0.2cm}
\centerline{\includegraphics[width=7cm]{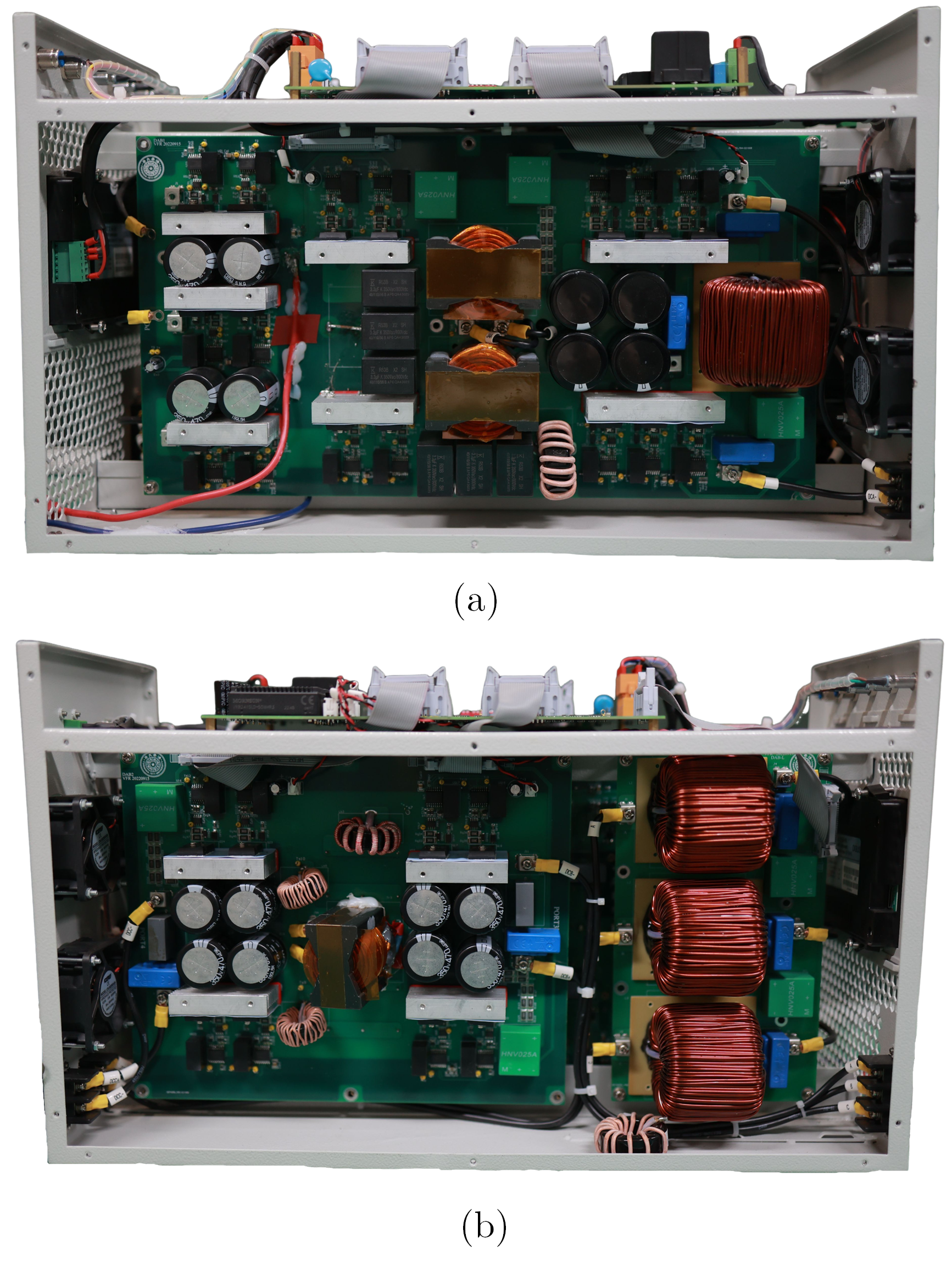}}
\caption{The experimental 4-port MAB prototype. (a) The front view. (b) The rear view.}
\label{fig:prototype}
\end{figure}

Simulations in MATLAB/Simulink regarding the above 4-port MAB converter have been conducted under the two working conditions, with each having the SPS control and the online ZVS design tested. Specifically, the steady-state high-frequency waveforms when $P_4$ = 2 kW under SPS and the full ZVS optimization are shown in Fig. \ref{fig:simulation_1}. In Fig. \ref{fig:simulation_1}(a), the SPS control leads to large RMS inductor currents in ports 2-4, and ports 3 and 4 lose ZVS operation. However, as shown in Fig. \ref{fig:simulation_1}(b), the online ZVS design achieves ZVS operation in all ports and significantly reduces the overall RMS inductor currents when compared to SPS control. The steady-state simulation shows that the online ZVS design has potentials to reduce switching and conducting losses, thus improving the overall efficiency of the converter. More $P_4$ points will be conducted in experiments to draw efficiency curves.

In addition, the dynamic powers between $P_4$ = 0.4 kW and $P_4$ = 2 kW under the online full ZVS design are recorded, as shown in Fig. \ref{fig:simulation_2}. Fig. \ref{fig:simulation_2}(a) shows that when $P_4$ increases from 0.4 kW to 2 kW, $P_2$ and $P_3$ remain stable and the whole dynamic process of tracking demanded $P_4$ ends within 1 ms. Another scenario where $P_4$ drops from 2 kW to 0.4 kW is shown in Fig. \ref{fig:simulation_2}(b). The process is even faster and smoother. Therefore, the simulations indicate that the proposed online full ZVS design has great dynamic responses towards load changes, which ensures reliable power tracking in real-time applications.

\begin{figure*}[ht]
\setlength{\abovecaptionskip}{-0.2cm}
\centerline{\includegraphics[width=18cm]{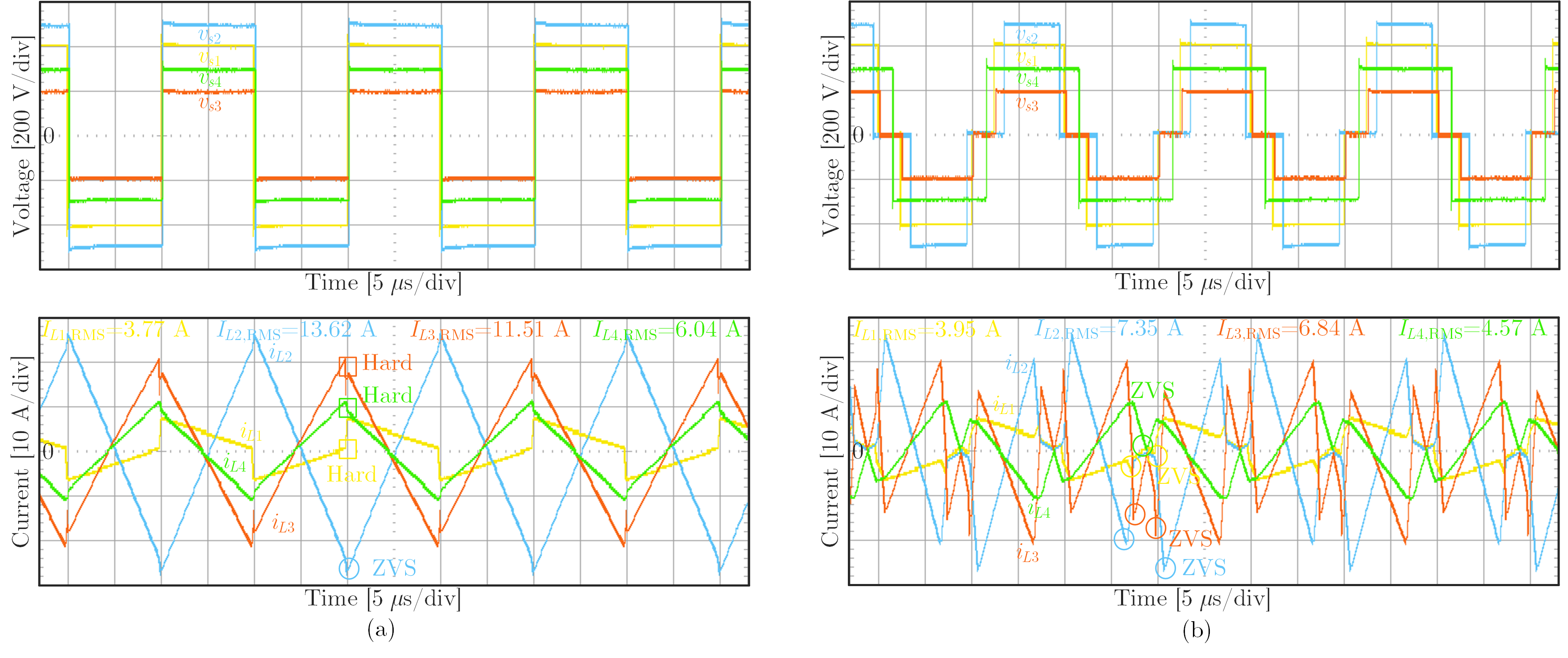}}
\caption{Experimental high-frequency waveforms when $P_4$=0.4 kW. (a) Under SPS control. (b) Under online ZVS design.}
\label{fig:experiment_1}
\end{figure*}

\begin{figure*}[ht]
\setlength{\abovecaptionskip}{-0.2cm}
\centerline{\includegraphics[width=18cm]{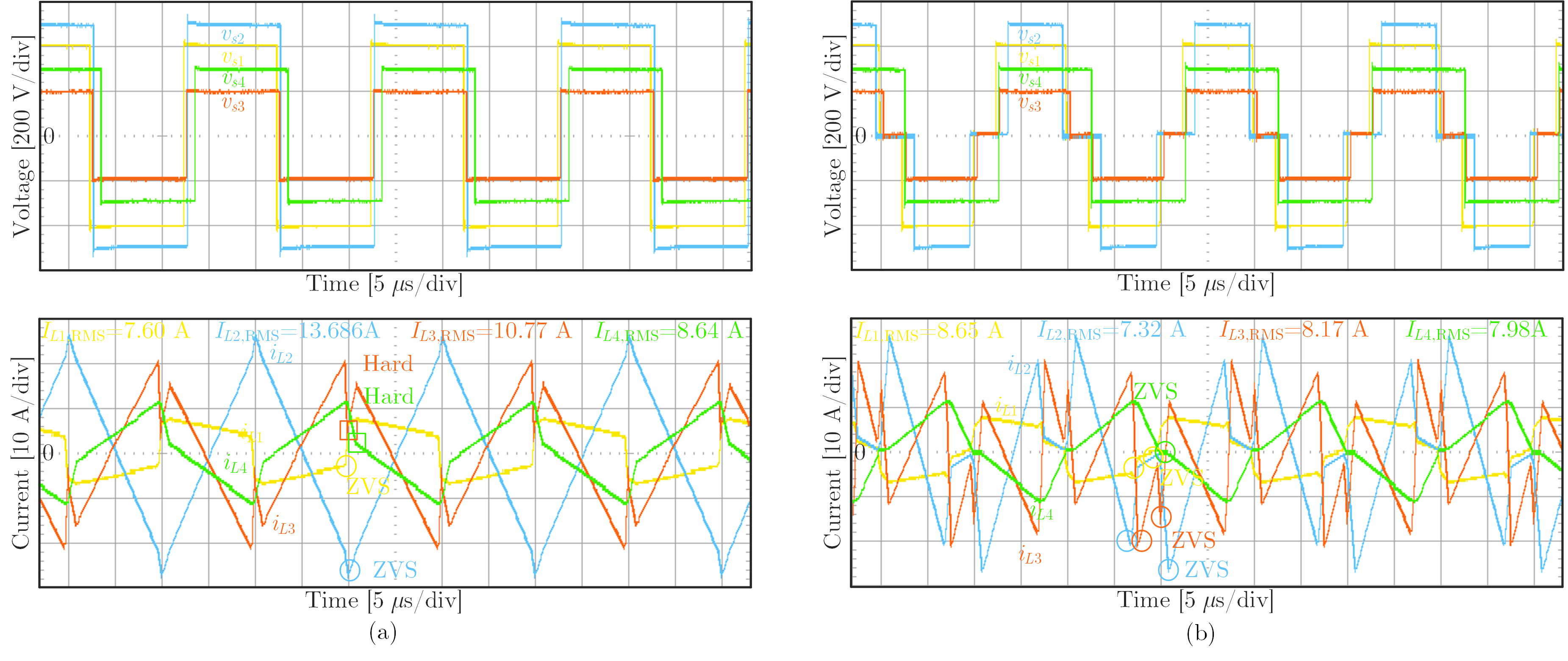}}
\caption{Experimental high-frequency waveforms when $P_4$=2 kW. (a) Under SPS control. (b) Under online ZVS design.}
\label{fig:experiment_2}
\end{figure*}

\subsection{Experimental Results}\label{subsec:experiment}
A 4-port MAB converter prototype with the same parameters has also been constructed for experimental verification, as shown in Fig. \ref{fig:prototype}. The experimental steady-state high-frequency waveforms when $P_4$ = 0.4 kW and $P_4$ = 2 kW are shown in Figs. \ref{fig:experiment_1} and \ref{fig:experiment_2}, respectively. 

As shown in Fig. \ref{fig:experiment_1}(a), only port 2 achieves ZVS operation and all other ports maintain hard switching under SPS control when $P_2$ = 0.4 kW, $P_3$ = 0.5 kW, and $P_4$ = 0.4 kW. This is because the voltage conversion ratio of port 2 is the largest and the converter is working under light load conditions. However, even with light load, the converter still has large RMS inductor currents due to the inherent backflow powers among the ports. Specifically, the RMS currents of all ports are $I_{L1,RMS}$ = 3.77 A, $I_{L2,RMS}$ = 13.62 A, $I_{L3,RMS}$ = 11.51 A, and $I_{L4,RMS}$ = 6.04 A. However, when the ZVS design is implemented, all ports retain ZVS operation, as shown in Fig. \ref{fig:experiment_1}(b), leading to significantly improved switching performance. It is notable that the waveform of $v_{s4}$ is two-level while other switched-node voltages are three-level ones. The design also reduces the backflow power and leads to RMS inductor currents of $I_{L1,RMS}$ = 3.95 A, $I_{L2,RMS}$ = 7.35 A, $I_{L3,RMS}$ = 6.84 A, and $I_{L4,RMS}$ = 4.57 A, which reduces 62.76\% of the total original RMS inductor currents in the SPS control and largely minimizes conduction losses.

\begin{figure*}[ht]
\setlength{\abovecaptionskip}{-0.2cm}
\centerline{\includegraphics[width=18cm]{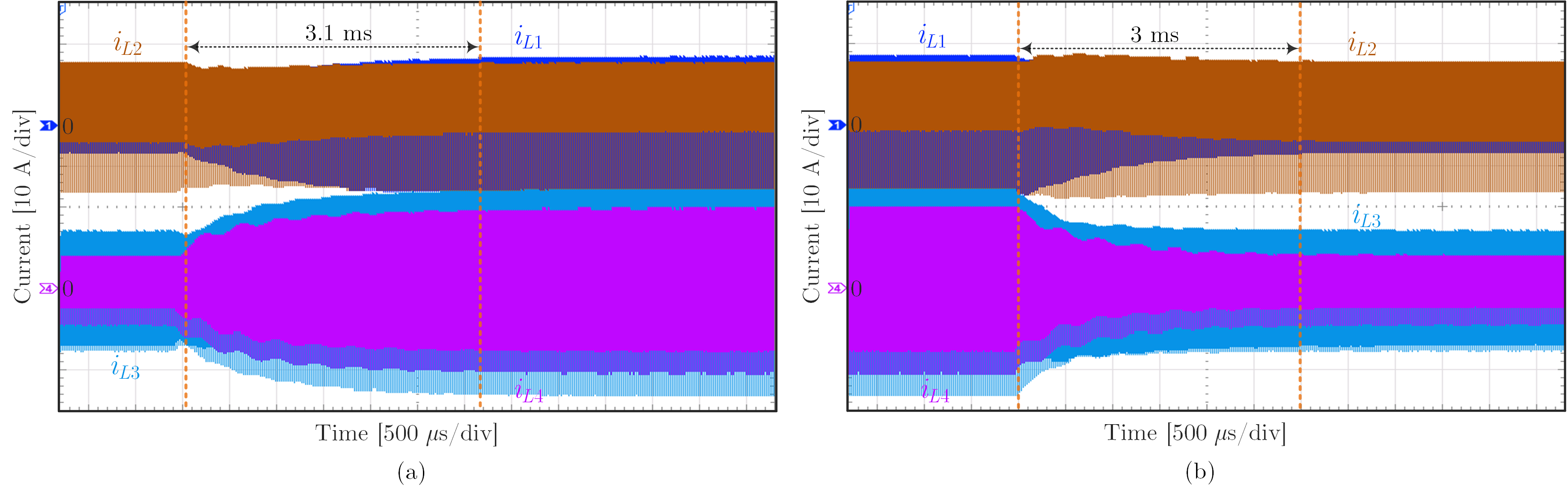}}
\caption{Experimental dynamic high-frequency inductor currents between $P_4$=0.4 kW and $P_4$=2 kW. (a) When $P_4$ changes from 0.4 kW to 2 kW. (b) When $P_4$ changes from 2 kW to 0.4 kW.}
\label{fig:experiment_3}
\end{figure*}

\begin{figure}[t]
\centerline{\includegraphics[width=9cm]{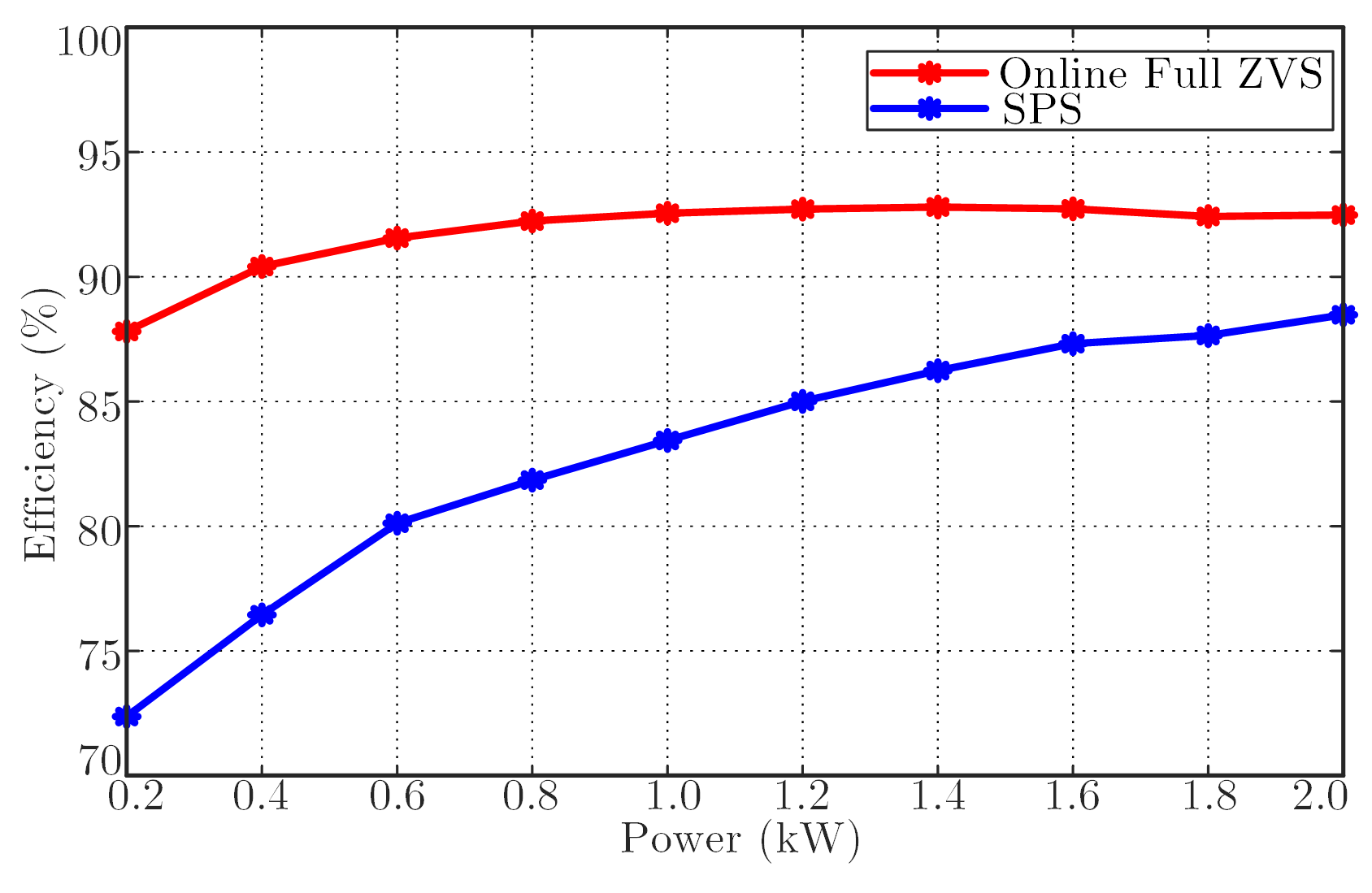}}
\caption{The efficiency curves under the SPS control and the proposed online full ZVS design within a wide operating range.}
\label{fig:efficiency}
\end{figure} 

The high-frequency waveforms of the converter under another working scenario are shown in Fig. \ref{fig:experiment_2} when $P_2$ = 0.4 kW, $P_3$ = 0.5 kW, and $P_4$ = 2 kW. The experimental waveforms are basically the same as the simulation waveforms shown in Fig. \ref{fig:simulation_1}. Similarly, ports 3 and 4 lose ZVS operation under SPS control and the overall RMS inductor currents are $I_{L1,RMS}$ = 7.60 A, $I_{L2,RMS}$ = 13.68 A, $I_{L3,RMS}$ = 10.77 A, and $I_{L4,RMS}$ = 8.64 A, as shown in Fig. \ref{fig:experiment_2}(a). The performance of the ZVS design under the same working condition is shown in Fig. \ref{fig:experiment_2}(b), where all ports also achieve ZVS operation. The RMS inductor currents under the ZVS design become $I_{L1,RMS}$ = 8.65 A, $I_{L2,RMS}$ = 7.32 A, $I_{L3,RMS}$ = 8.17 A, and $I_{L4,RMS}$ = 7.98 A, which is only 59.43\% of the RMS values under SPS control. Therefore, the online ZVS design also improves the ZVS performance and reduces the overall RMS inductor currents under the heavy load condition for port 4. Thus, the significant improvement verifies the effectiveness of the proposed online full ZVS design.

The dynamic response of the full ZVS design has also been tested experimentally. The transient internal high-frequency inductor currents between $P_4$ = 0.4 kW and 2 kW are shown in Fig. \ref{fig:experiment_3}. As shown in Fig. \ref{fig:experiment_3}(a), when $P_4$ changes from 0.4 kW to 2 kW, it takes nearly 3.1 ms for the inductor currents to stabilize, while the dynamic response time is about 3 ms when $P_4$ drops from 2 kW to 0.4 kW, as shown in Fig. \ref{fig:experiment_3}(b). The tests indicate that an average response time of 3 ms is obtained during the power-up-and-down transitions, which is at the same level as the traditional SPS control and can be improved by optimizing the PI parameters if needed. Most importantly, the results show that the proposed ZVS design is able to handle load changes without preprocessed look-up tables that are usually needed in state-of-the-art optimization strategies, thus achieving online full ZVS optimization and ensuring fast and secure operation during real-time implementation.

Finally, efficiency tests have been performed for the SPS control and the proposed ZVS design within a wide operating range. The power handled by port 4 has been set from 0.2 kW to 2 kW with a step size of 0.2 kW. The efficiency curves are shown in Fig. \ref{fig:efficiency}. Specifically, a maximum efficiency of 92.8\% is obtained at $P_4$ = 1.4 kW even under the extreme voltage conversion ratios and inductance coefficients, and a maximum efficiency improvement of 15.44\% has been achieved at $P_4$ = 0.2 kW when compared to the SPS control. The efficiency testes show that in addition to maintaining great dynamic responses, the proposed online ZVS design has significantly reduced the overall power losses and improved the efficiency of the converter.

\section{Conclusion}\label{sec:conclusion}
This paper proposes an online full ZVS optimization design for generalized MAB converters in MV PET. By establishing full ZVS equations and simplifying the solution, an online control strategy utilizing all inner phase-shift ratios has been proposed with only sampled DC voltages needed. Simulation and experimental results show that compared to conventional single phase-shift control, the proposed method achieves better ZVS operation and less RMS currents while maintaining great dynamic performance. The proposed online full ZVS design significantly improves the system efficiency and can be implemented in real-time applications within a wide operating range.

\vspace{12pt}
\color{red}

\end{document}